\documentclass{article}

\usepackage[preprint]{neurips_2020}

\usepackage[utf8]{inputenc} 
\usepackage[T1]{fontenc}    
\usepackage{hyperref}       
\usepackage{url}            
\usepackage{booktabs}       
\usepackage{amsfonts}       
\usepackage{nicefrac}       
\usepackage{microtype}      
\usepackage{graphicx}
\usepackage{amsmath}

\usepackage{natbib}

\usepackage{hyperref}
\hypersetup{
    colorlinks=true,
    linkcolor=blue,
    filecolor=black,      
    urlcolor=blue,
    citecolor=blue
}

\title{RRF102: Meeting the TREC-COVID Challenge \\ with a 100+ Runs Ensemble}
\author{%
 Michael Bendersky\textsuperscript{1}, Honglei Zhuang\textsuperscript{1}, Ji Ma\textsuperscript{1},Shuguang Han\textsuperscript{2}\thanks{Work done while at Google Research.}, Keith Hall\textsuperscript{1}, Ryan McDonald\textsuperscript{1} \\
 (1) Google Research (2) Alibaba Group \\
(1) \texttt{\{bemike,hlz,maji,kbhall,ryanmcd\}@google.com} (2) \texttt{hanshuguang@gmail.com} \\
}

\newcommand{\corpus}{\ensuremath{\mathcal{C}}}
\newcommand{\doc}{\ensuremath{d}}
\newcommand{\runPool}{\ensuremath{\mathcal{R}}}
\newcommand{\system}{\ensuremath{S}}
\newcommand{\run}{\ensuremath{r}}
\newcommand{\permutation}[1]{\ensuremath{\pi_{#1}}}
\newcommand{\rank}[2]{\ensuremath{\permutation{#1}({#2})}}
\newcommand{\score}[2]{\ensuremath{\text{sc}_{#1}({#2})}}
\newcommand{\rrf}{\ensuremath{\text{rrf}}}
\newcommand{\hrrf}{\ensuremath{\text{hrrf}}}
\newcommand{\Bert}{\ensuremath{\text{BERT}}}
\newcommand{\Roberta}{\ensuremath{\text{RoBERTa}}}
\newcommand{\Electra}{\ensuremath{\text{ELECTRA}}}
\newcommand{\hwrrf}{\ensuremath{\text{h$_{w}$rrf}}}
\newcommand{\teamname}[1]{\texttt{#1}}

\def\be{{\mathbf{e}}}
\def\bW{{\mathbf{W}}}
\def\CLS{{\texttt{[CLS]}}}
\def\SEP{{\texttt{[SEP]}}}

\begin{document}

\maketitle

\begin{abstract}
In this paper, we report the results of our participation in the TREC-COVID challenge. To meet the challenge of building a search engine for rapidly evolving biomedical collection, we propose a simple yet effective weighted hierarchical rank fusion approach, that ensembles together 102 runs from (a) lexical and semantic retrieval systems, (b) pre-trained and fine-tuned BERT rankers, and (c) relevance feedback runs. Our ablation studies demonstrate the contributions of each of these systems to the overall ensemble. The submitted ensemble runs achieved state-of-the-art performance in rounds 4 and 5 of the TREC-COVID challenge.
\end{abstract}

\section{Introduction}

In this paper, we analyze the participation of our team -- \teamname{unique\_ptr} -- in the TREC-COVID challenge organized by the Allen Institute for Artificial Intelligence (AI2), the National Institute of Standards and Technology (NIST), the National Library of Medicine (NLM), Oregon Health and Science University (OHSU), and the University of Texas Health Science Center at Houston (UTHealth)\footnote{\scriptsize{\url{https://ir.nist.gov/covidSubmit/}}}.  The TREC-COVID challenge followed the TREC model for building test collections through community participation; the submissions from the different teams were pooled to create a reusable test collection to encourage future research in systems for information retrieval from scientific literature. The CORD-19 Research Dataset\footnote{\scriptsize{\url{https://www.semanticscholar.org/cord19}}} was used as a target retrieval corpus.

\begin{figure}
\centering
\begin{tabular}{cc}
\includegraphics[width=0.47\columnwidth]{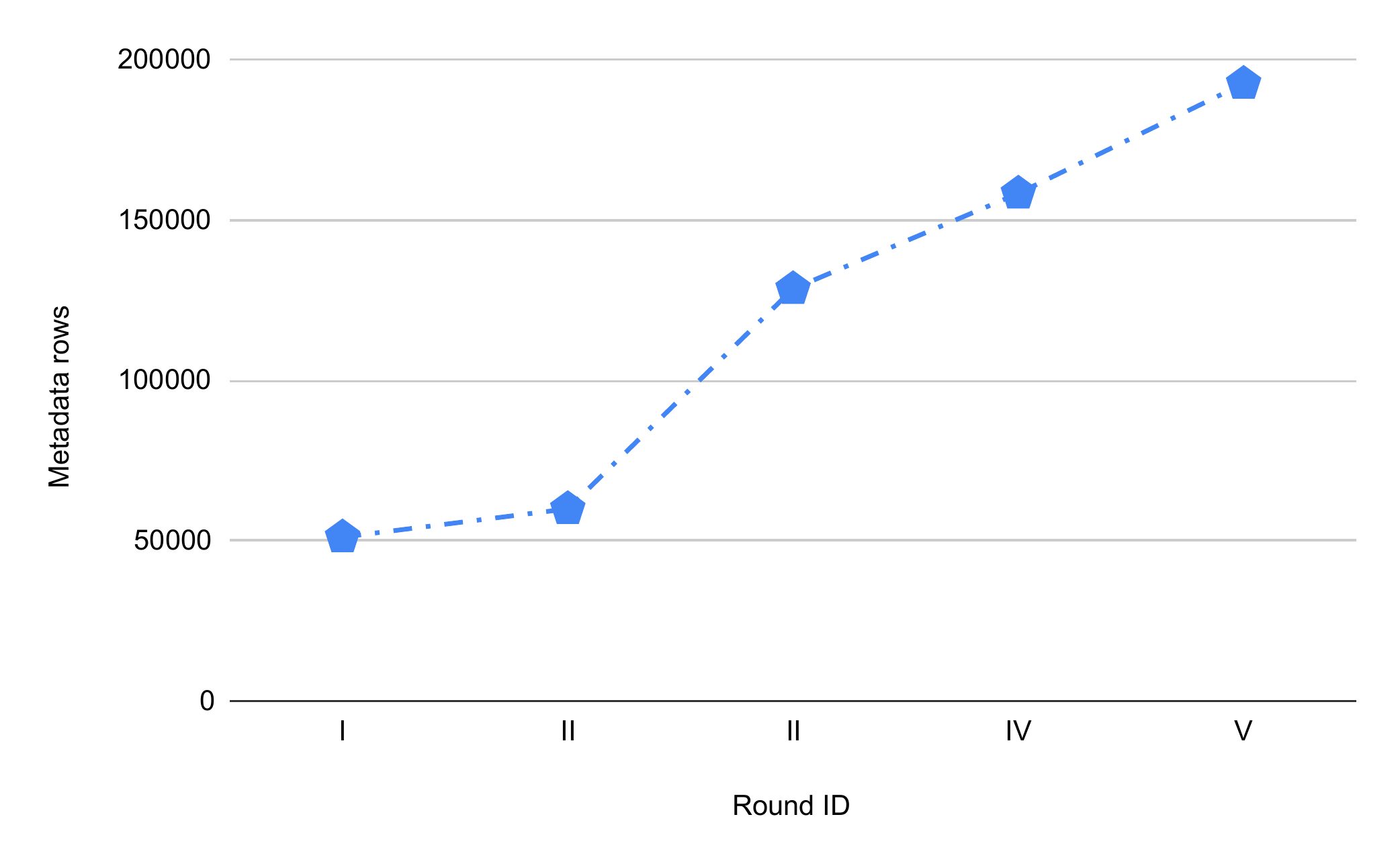}  &  
\includegraphics[width=0.47\columnwidth]{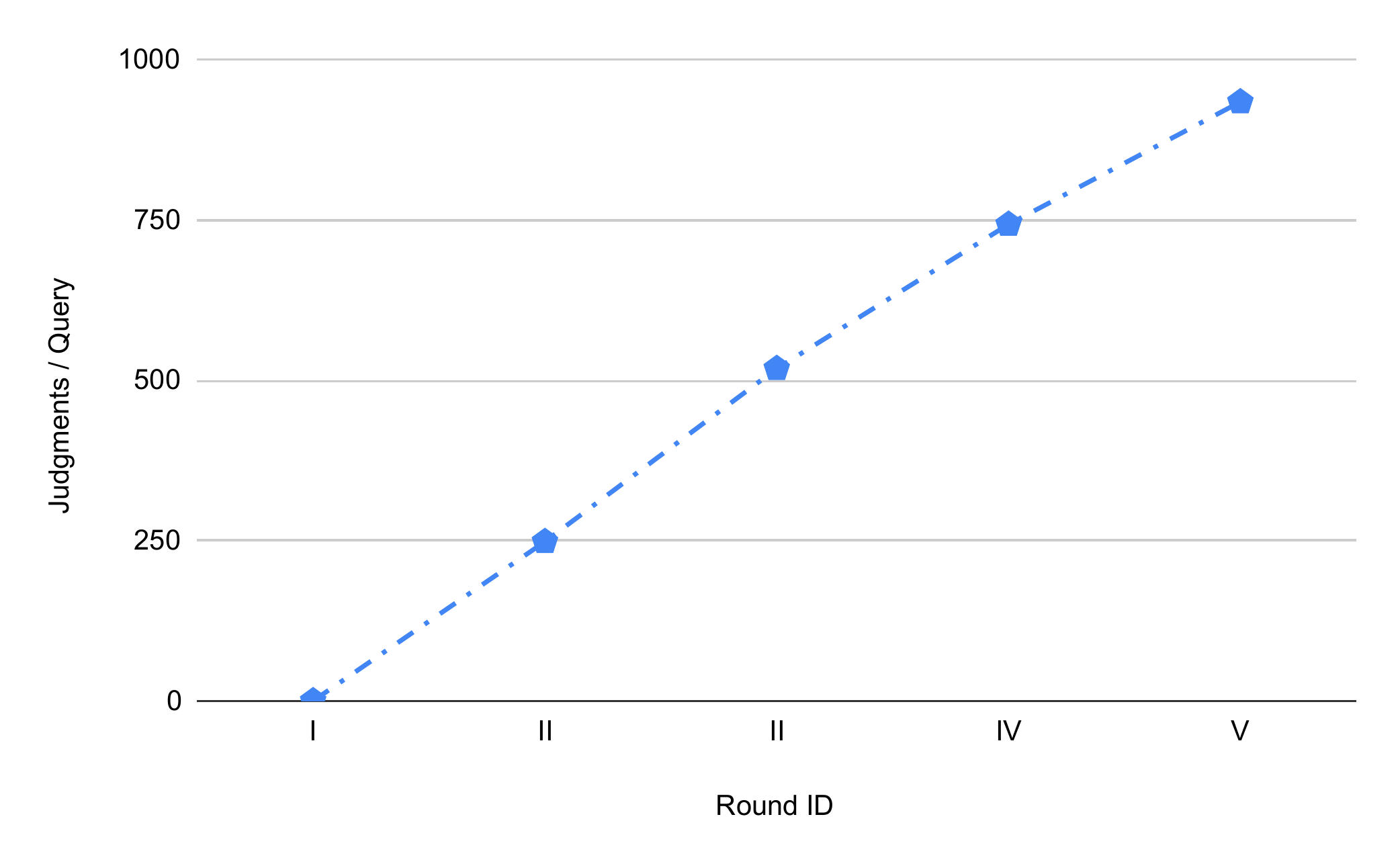} \\
(a) & (b) \\
\end{tabular}
\caption{Plots that demonstrate the evolution of the CORD-19 corpus as the TREC-COVID rounds progressed. (a) Number of documents for which any metadata was available in each round; (b) Available relevance judgments per query prior to each round.}
\label{fig:corpus-growth}
\end{figure}

The challenge was organized as a series of five rounds, where participants could choose to skip any round. Each round was associated with a set of structured search topics for which relevant documents needed to be retrieved (see Figure~\ref{fig:topic-example} for an example of a topic). While the majority of the topics repeated across rounds with only five new topics added per-round, the CORD-19 dataset itself grew significantly during the time that the challenge took place (see Figure \ref{fig:corpus-growth}(a)), and only the new relevance assessments (by human annotators with biomedical expertise) from the round were used to score the submissions. Therefore, strong performance in one round did not guarantee success in future rounds. \teamname{unique\_ptr} had participated in all five rounds, however the techniques described in this paper only solidified in rounds 4 and 5, where our team achieved the best performance among 72 and 126 runs, respectively, on the majority of the evaluation metrics.

Early on in the competition, we realized that due to the rapid evolution of the corpus, it is unlikely that the ``winner takes all'' approach will dominate, with a single method leading the challenge in all rounds. Therefore, instead, we turned our attention to an ensemble approach, that would be able to adapt to the rapidly evolving CORD-19 content, and would be able to leverage the growing pool of relevance judgements for each query (Figure \ref{fig:corpus-growth}(b)). After several less successful (but highly educational) attempts in Rounds 1 -- 3, we have zeroed in on the ensemble approach described in this paper. 

Our approach combines runs from lexical and semantic retrieval systems, as well as pre-trained BERT rankers to achieve a dual effect of high recall of relevant retrieved documents, and high precision at the top ranks. It can also make use of existing relevance judgements, both in the retrieval stage via relevance feedback, as well as in the ranking stage, via BERT model fine-tuning.

To join all of these disparate retrieval and ranking components together we propose a simple but effective \emph{weighted hierarchical rank fusion} technique. Our final submission -- codenamed \teamname{RRF102} -- ensembles together $102$ different retrieval and ranking runs using this technique.  \teamname{RRF102} significantly and consistently outperforms other alternatives both in our ablation studies, as well as on the official \href{https://docs.google.com/spreadsheets/d/1n1IJ6gkZQh3lyjJFQZPOm2WL5Ct3GFSWa7ImqBhcvn4/edit?usp=sharing}{Round 5 leaderboard}.

\section{Related Work}
Readers who are interested in the further details on the TREC-COVID challenge are encouraged to refer to an excellent overview by~\citet{Voorhees+al:2020}. There were also other publications by the participating teams, most of which can be found on the \href{https://ir.nist.gov/covidSubmit/bib.html}{TREC-COVID Bibliography page}. We do not claim novelty for most of the ideas presented in this paper. Many of them were discussed and utilized by other challenge participants, like relevance feedback~\citep{Zhang+al:2020} or using the MS-MARCO dataset for training BERT ranking model~\citep{MacAvaney+al:2020}.  Our main contribution in this work is careful evaluation of the various retrieval and ranking systems, as well as a robust ensembling mechanism that combines lexical and semantic retrieval with multiple rankers.

\begin{figure}
\centering
\small{
\begin{verbatim}
<topic number="49">
  <query> post-infection COVID-19 immunity </query>
  <question> do individuals who recover from COVID-19 show sufficient
             immune response, including antibody levels and T-cell
             mediated immunity, to prevent re-infection?
  </question>
  <narrative> There is concern about re-infection for COVID-19, so this
              topic is looking for studies suggesting post-infection immunity,
              including post-infection antibody levels (over time) and
              evidence for individuals who have been infected more than once.
  </narrative>
</topic>
\end{verbatim}}
\caption{Example of a topic used in TREC-COVID challenge.}
\label{fig:topic-example}
\end{figure}

\section{Ensemble Construction}
\label{sec:ensemble}

Ensemble models have repeatedly been placed at the top positions of recommendation and ranking competitions. For instance, \citet{Bell+Koren:2007}, who were among the winners of the Netflix prize noted that: ``\emph{it was important to utilize a variety of models that complement the shortcomings of each other}''. \citet{Burges+al:2011} -- the winners of the Yahoo! Learning to Rank challenge -- used a linear combination of 12 ranking models, combining LambdaMART boosted decision trees, neural nets and logistic regression with various loss functions. \citet{Han+al:2020} used an ensemble of 15 BERT, RoBERTa and ELECTRA models to achieve top best performance in the MS-MARCO passage reranking task. Therefore, following these success stories, we also focused on exploring ensemble models for our TREC-COVID challenge submission.

The most common application of the ensembling method (e.g., in a classification or a regression task) consists of two main stages: first, we develop a pool of base learners, and then combine them together to form an aggregate prediction~\citep{Hastie+al:2009}. Since each of these two stages may require some fitting to the training data, it is important to ensure that the overall ensemble is not so complex that it is overfitted, and does not generalize to unseen data. Therefore, in most cases, ensembles are implemented using a pool of simple diverse base learners usually combined via (weighted) averaging of their predictions.

The setting we face in the TREC-COVID challenge is somewhat more involved than the ensemble setting presented above. As is common in information retrieval, there are two stages to generating the optimal ranked list. First, we need to \emph{retrieve} an initial set of candidates that potentially match the topic. The success at this stage is measured by a metric like \emph{recall@K} (where $K \approx O(1000)$). Then, we need to apply \emph{ranking} to this set, to achieve the most optimal ordering. The success in the ranking stage will be measured by a metric like \emph{nDCG@K} (where $K \approx O(10))$. Mean average precision (\emph{MAP}) is often used as an effective metric for measuring the joint effect of the two stages (both high recall \emph{and} precision) .

Many of the successful submissions at the first rounds of the TREC-COVID challenge fixed the first stage (retrieval) to be a simple lexical retrieval algorithm, e.g., BM25, and instead focused on the re-ranking stage~\citep{MacAvaney+al:2020}. While reasonable (and indeed effective at achieving high \emph{nDCG@10} performance), in our opinion, this approach may limit the real-world applicability of the resulting algorithms, since high recall is of importance in the medical domain, e.g., for summarizing all available evidence regarding a certain treatment or symptom~\citep{Kanoulas+al:2017}.

Therefore, we have deployed a two-pronged ensembling approach in our submission. On one hand, we use a combination of different retrieval mechanisms to obtain a comprehensive document set to increase recall. On the other hand, we also apply multiple BERT-based rankers to this set, which was found to be beneficial for high precision at top ranks in prior studies~\citep{Han+al:2020}. As we demonstrate in Section~\ref{sec:exp}, the resulting ensemble achieves the best of both worlds: state-of-the-art performance on a wide range of both precision and recall metrics. 

How to optimally construct such two-stage retrieval and ranking ensembles remains an interesting open research question that we will undoubtedly revisit in the future. However, for the purpose of TREC-COVID challenge, using reciprocal rank fusion (RRF)~\citep{Cormack+al:2009} as a foundational building block, results in robust and effective ensembles. One important advantage of RRF is that, as its name suggests, it only requires access to document ranks, and thus can accommodate heterogeneous candidate sets with differing score ranges. However, in its most basic form, RRF can result in sub-optimal performance, which we address by proposing a simple hierarchical variant of this method.

\subsection{Notation}

We first introduce some notation that will be useful in the exposition of our method, which follows next. We are given a document corpus \corpus, from which the candidate documents are drawn. We are also given a set of runs \runPool, wherein each run $\run \in \runPool$ induces a permutation \permutation{\run} over a subset of documents $\{d\} \subset \corpus$. With a slight abuse of notation, we denote the rank of document $d$ in run $r$ as \rank{\run}{\doc}, and its respective score as \score{\run}{\doc}. Each run $r$ is generated by some system \system, and therefore the entire run pool \runPool~can be divided into non-overlapping pool subsets $\runPool_\system$. Note that each system \system~can generate multiple runs, e.g., through variation in parameters or inputs. The concrete implementations of systems and runs are discussed in detail in Section \ref{sec:systems}. In the remainder of this section, we discuss how these runs are ensembled to form our final submission.

\subsection{Reciprocal Rank Fusion}
\label{sec:ensemble:rrf}

Following the formulation originally proposed by \citet{Cormack+al:2009}, reciprocal rank fusion (RRF) sorts the documents according to a simple scoring formula, where document score is defined as a sum of reciprocal ranks of the document across the runs:

\begin{equation}
    \score{\rrf}{\doc} = 
    \begin{cases}
    \sum_{\run \in \runPool} \frac{1}{k + \rank{\run}{\doc}},& \text{if } d \in \permutation{\run} \\
    0,              & \text{otherwise.}
    \label{eq:rrf-flat}
\end{cases}
\end{equation}

We fix $k = 60$, following the original paper. $k$ is a constant that mitigates the impact of high rankings by outlier systems. Sorting all the documents where $\score{\rrf}{\doc} > 0$ in a descending order of \score{\rrf}{\doc} produces an RRF run \permutation{\rrf}.

\subsection{Hierarchical Reciprocal Rank Fusion}
\label{sec:ensemble:hrrf}

As noted above, our runs from heterogeneous set of systems and the number of runs across systems may vary quite dramatically (see more on that in Section~\ref{sec:systems}). Therefore, rankings in the \permutation{\rrf}~run may be dominated by the system that has the most runs. To mitigate this effect, and to ensure that no system is over-represented in the final fusion run, we propose a simple approach based on a \emph{hierarchical} application of rank fusion.

First we divide our run pool \runPool, into sub-pools, $\runPool_\system$, each corresponding to runs produced by system \system. Obviously, it is possible to divide \runPool~into sub-pools beyond system pools, but we stick to this simple mechanism in our submissions, as it is quite logical, and empirically effective.

For each pool $\runPool_\system$, we produce a single run \permutation{\rrf_\system}, such that 

\begin{equation*}
    \score{\rrf_\system}{d} = 
    \begin{cases}
    \sum_{\run \in \runPool_\system} \frac{1}{k + \rank{\run}{\doc}},& \text{if } d \in \permutation{\run} \\
    0,              & \text{otherwise.}
    \label{eq:rrf-system}
\end{cases}
\end{equation*}

We then recursively apply RRF to all the \permutation{\rrf_\system} runs, resulting in the final hierarchical rank fusion run \permutation{\hrrf}:

\begin{equation}
    \score{\hrrf}{d} = 
    \begin{cases}
    \sum_{\runPool_\system \subset \runPool} \frac{1}{k + \rank{\rrf_\system}{\doc}},& \text{if } d \in \permutation{\rrf_\system} \\
    0,              & \text{otherwise.}
    \label{eq:h-rrf}
\end{cases}
\end{equation}

\subsection{Weighted Hierarchical Reciprocal Rank Fusion}
\label{sec:ensemble:hwrrf}

Since it is likely that not all systems are of equal quality, intuitively it makes sense to weight their contributions, resulting in a variant of the hierarchical rank fusion \permutation{\hwrrf}:

\begin{equation}
    \score{\hwrrf}{d} = 
    \begin{cases}
    \sum_{\runPool_\system \subset \runPool} \frac{w_S}{k + \rank{\rrf_\system}{\doc}},& \text{if } d \in \permutation{\rrf_\system} \\
    0,              & \text{otherwise.}
    \label{eq:hw-rrf}
\end{cases}
\end{equation}

Given some training data, it is possible to make the weights $w_\system$ learnable. However, given the paucity of training data, we were somewhat apprehensive of overfitting, and used a simple heuristic instead. We set $w_\system = 2$ for any systems that rely on prior relevance judgments, and $w_\system = 1$ for all other systems. This heuristic has the advantage of reflecting the intuition that the systems that had access to human labels are more trustworthy, without explicitly using any human labels to estimate the level of this trust. 

\begin{figure}[t]
\centering
\includegraphics[width=0.95\columnwidth]{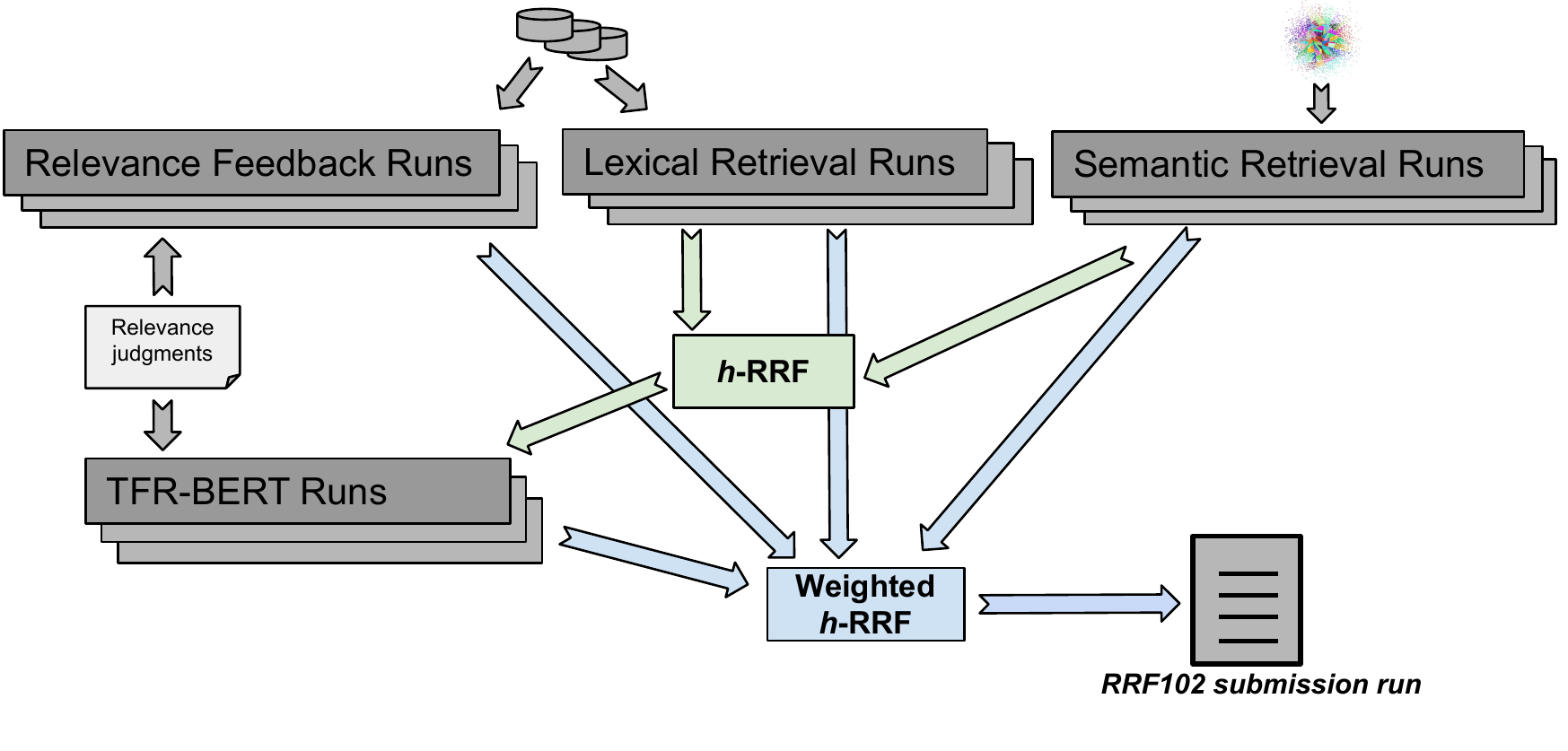} 
\caption{Schematic diagram of the weighted hierarchical rank fusion ensemble.}
\label{fig:system-schematic}
\end{figure}

\section{Detailed Overview of Systems and Runs}
\label{sec:systems}

Thus far, we described hierarchical reciprocal rank fusion ($h$-RRF), the general ensembling framework within which we operated in our submissions. In this section, we provide a detailed exposition of the systems that were ensembled using $h$-RRF, and elaborate on the runs that were produced using these systems. In general, as discussed in the beginning of the previous section, we gave preference to simple, replicable systems that require as little training data as possible.

\begin{table*}[]
    \centering
    \begin{tabular}{|l|l|l|c|}
    \hline
     & System & Type & \# Runs Produced \\
    \hline
    1 & Terrier & Lexical Retrieval & 14 \\
    2 & Anserini & Lexical Retrieval & 12 \\
    3 & Dual Encoder & Semantic Retrieval & 24 \\
    4 & Terrier & Relevance Feedback & 2 \\
    5 & Anserini & Relevance Feedback & 2\\
    6 & MS-Marco BERT & TFR-BERT & 30 \\
    7 & Finetuned BERT & TFR-BERT & 18\\
    \hline
    \multicolumn{3}{|c}{} & \textbf{Total: 102} \\
    \hline
    \end{tabular}
    \caption{Summary of retrieval and ranking systems used, and the runs produced by each system.}
    \label{tab:all-runs}
\end{table*}

Figure~\ref{fig:system-schematic} provides a schematic overview of our overall ensembling flow. First, we use lexical and semantic retrieval from either inverted indices or a k-nearest neighbor database, respectively, to retrieve a set of candidates, ranked by a simple match score (e.g., BM25 or vector dot product). The candidates from the runs generated by these retrieval systems are fed into a hierarchical rank fusion (as shown in Equation~\ref{eq:h-rrf}), and its output is re-ranked using multiple Tensorflow Ranking BERT models~\citep{Han+al:2020}. In addition, we perform several runs of a standard relevance feedback-based retrieval. 

Finally, the outputs of these four systems (lexical and semantic retrieval, relevance feedback and TFR-BERT) are all fused used weighted $h$-RRF (Equation~\ref{eq:hw-rrf}). This constitutes our best final submission run; we refer to it as \texttt{RRF102}, with the name indicating the total number of runs being ensembled. A summary of these runs and the systems that produced them is provided in Table~\ref{tab:all-runs}. We use the remainder of this section to describe them in more detail.

\subsection{Lexical Retrieval Systems}

We used two popular open source search engines, Terrier~\citep{Terrier:2005} and Anserini~\citep{Anserini:2017} to generate our lexical retrieval runs. Terrier was elected based on its excellent documentation, expressive query language, and its ability to implement common retrieval algorithms via configuration. In the case of Anserini, we used the runs kindly published by the \teamname{covidex} team~\citep{Zhang+al:2020}. These runs were also used by many of the competing teams, thus providing a natural benchmark for evaluating the performance of the other systems in the ensemble.

\subsubsection{Terrier}
For the Terrier lexical retrieval system we generate multiple retrieval runs, each using a different representation of a subset of topic fields:
\begin{itemize}
    \item Bag-of-words representation of the \texttt{query} field
    \item DFR-based dependence model~\citep{Peng+al:2007} representation of the \texttt{query} field
    \item Bag-of-words representation of the \texttt{question} field
    \item DFR-based dependence model representation of the \texttt{question} field
    \item Bag-of-words representation of the concatenation of \texttt{query} and \texttt{question} fields
    \item Bag-of-words representation of the concatenation of \texttt{question} and \texttt{narrative} fields
    \item Bag-of-words representation of the concatenation of \texttt{query} and \texttt{question} fields, expanded with 10 most informative terms appearing in the top documents.
\end{itemize}

For each of these representations, we apply the resulting queries to both abstract and full-text indices. In all the runs, unless specified otherwise, we use the default Terrier settings. Overall, this results in 14 Terrier lexical retrieval runs.

\subsubsection{Anserini}
For Anserini, we do not conduct any runs ourselves, but rather use the runs provided by the \teamname{covidex} team. Specifically we use the following combinations of indices and topic fields:
\begin{itemize}
    \item \texttt{abstract} AND \texttt{query+question}
    \item \texttt{abstract} AND \texttt{UDel-qgen}
    \item \texttt{full-text} AND \texttt{query+question}
    \item \texttt{full-text} AND \texttt{UDel-qgen}
    \item \texttt{paragraph} AND \texttt{query+question}
    \item \texttt{paragraph} AND \texttt{UDel-qgen}
\end{itemize}

We use these combinations both for regular\footnote{\scriptsize{\url{https://github.com/castorini/anserini/blob/master/docs/experiments-covid.md}}} and doc2query expanded\footnote{\scriptsize{\url{https://github.com/castorini/anserini/blob/master/docs/experiments-covid-doc2query.md}}} Anserini indices, resulting in 12 runs. Note that we do not use any of the published Anserini rank fusion runs, as we rely on our own implementation of the hierarchical rank fusion. 

\subsection{Relevance Feedback Systems}
\label{sec:systems:rf}

As in each of the TREC-COVID challenge rounds the majority of the existing topics were being reused, past participants found relevance feedback to be beneficial for obtaining effective submissions. As an example, \texttt{UIowaS} team had achieved consistently high ranks across multiple rounds using a simple Borda fusion of multiple Terrier runs with BM25 weighting and relevance feedback. 

Inspired by this simple yet effective approach we implement two relevance feedback runs using our Terrier system with abstract index:
\begin{itemize}
    \item Relevance feedback run with \texttt{query+question} field expanded by $300$ terms from $10$ highest ranked relevant documents.
    \item Relevance feedback run with \texttt{query+question} field expanded by $1,000$ terms from $30$ highest ranked relevant documents.    
\end{itemize}

In addition, we use two published Anserini relevance feedback runs using both regular and doc2query expanded abstract indices. Overall, this results in four relevance feedback runs used in our submission.

\subsection{Semantic Retrieval System}
\label{sec:systems:semantic}

\subsubsection{Neural retrieval model based on BERT}
\label{sec:de}
The neural retrieval model belongs to the family of \textit{relevance-based dense retrieval} or dual encoder models that encodes pairs of items in dense subspaces \citep{palangi2016deep}.  
In particular, our encoders are based on BERT \citep{bert}, which takes a query(or document) as input, and then projects the \CLS token representation down to a 768 dimensional vector as the embedding of that query (or document).  
We then compute the relevance score as vector dot product between the query and document embedding.
We share parameters between query and document encoder, so called Siamese networks -- as we found this greatly increased performance while reducing parameters. 

We train our dual-encoder models using softmax cross-entropy loss together with in-batch negatives, i.e.,
given a query in a batch of (query, relevant-passage) pairs, passages from other pairs are considered irrelevant for that query. In-batch negatives has been widely adopted in training neural network based retrieval models as it enables efficient training via computation sharing \citep{GillickE2E,dense-passage-qa}.

For serving, we first run the encoder over every passage offline to create a distributed lookup-table as a backend. 
At inference, we only need to run the encoder on the input query.
The query encoding is used to perform nearest neighbour search against the passage encodings in the backend.  
Since the total number of passages is in the order of millions and each passage is projected to a 768 dimensional vector, we use distributed brute-force search for exact inference instead of approximate nearest neighbour search \citep{liu2011hashing,JDH17}.

\subsubsection{Synthetic question generation}

One critical ingredient for training deep neural models is the abundant training data.
However, such resource is not always available, especially on specialized domains such as biomedical domain. 
To handle the data scarcity issue, we adopt the data augmentation approach proposed by \citet{Ma2020ZeroshotNR}, which
automatically generates synthetic questions on the target domain documents.
In particular, a transformer-based \citep{transformer} encoder-decoder generation model is trained to generate questions specific to a given passage.  
On completion, we apply the question generator on abstracts of PubMed/MEDLINE articles.  
This generates roughly 166 million (synthetic question, abstract) pairs for training our dual encoder model.


\subsubsection{Hybrid retrieval system}

Although dual encoder models are good at capturing semantic similarity, e.g., ``Theresa May'' and ``Prime Minister'' \citep{cross_domain_adv}, we observe lexical matching consistently poses a challenge for the dual encoder model.  
To mitigate the issue, we build a hybrid retrieval system by combining the dual encoder model with BM25 model,
exploiting the strength of BM25 in term matching.\footnote{Note that while the lexical match issue may be somewhat mitigated by the lexical retrieval systems in the ensemble, we did find combining with BM25 at a system level helpful in our investigations, as it provides more diversity in the runs to the final ensemble.}
Note that lexical retrieval systems like BM25 can be viewed as vector dot-product with nearest neighbor search.
Formally, let $\textbf{q}^{\text{bm25}} \in [0,1]^{|V|}$ be a $|V|$-dimensional binary encoding of a query $q$, i.e., $\textbf{q}^{\text{bm25}}[i]$ is 1 if the i-th entry of vocabulary $V$ is in $q$, 0 otherwise. 
Furthermore, let $\textbf{d}^{\text{bm25}} \in \mathbb{R}^{|V|}$ be a sparse real-valued vector where,
\begin{equation}
\textbf{d}_i^\text{bm25} = 
 \frac{\text{IDF}(d_i)* \text{cnt}(d_i, d)*(k + 1)}{\text{cnt}(d_i, d) + k*(1 - b + b * \frac{m}{m_\text{avg}})}. \nonumber
\end{equation}
We can see that,
\begin{equation}
    \text{BM25}(q, d) = \langle \textbf{q}^\text{bm25}, \textbf{d}^\text{bm25} \rangle \nonumber
\end{equation}
Here $\langle, \rangle$ denote vector dot-product.  This view gives rise to a simple hybrid model:
\begin{align}
    \nonumber
    \text{sim}(\textbf{q}^\text{hyb}, \textbf{d}^\text{hyb}) &= \langle \textbf{q}^\text{hyb}, \textbf{d}^\text{hyb} \rangle \\ \nonumber
    &= \langle [\lambda\textbf{q}^\text{nn},\textbf{q}^\text{bm25}], [\textbf{d}^\text{nn}, \textbf{d}^\text{bm25}] \rangle \\ \nonumber
    &= \lambda \langle \textbf{q}^\text{nn}, \textbf{d}^\text{nn} \rangle + \langle \textbf{q}^\text{bm25}, \textbf{d}^\text{bm25} \rangle,
\end{align}
where $\textbf{q}^\text{nn}$ and $\textbf{d}^\text{nn}$ denote query and document embedding from the dual encoder model, respectively. 
$\lambda$ is a hyper-parameter that controls the weight of the dual encoder system.

We use this hybrid system to generate multiple runs, based on different topic and index configurations:
\begin{itemize}
    \item \texttt{abstract} AND \texttt{query+question+narrative}
    \item \texttt{full-text} AND \texttt{query+question+narrative}
    \item \texttt{full-text} AND \texttt{query+question}
    \item \texttt{full-text} AND \texttt{query}
\end{itemize}

For each the above configuration, we also try different $\lambda$ values within $\{1, 5, 10, 15, 20, 30\}$. This results in $24$ overall dual encoder runs.

\subsubsection{Implementation Details}

Both the encoder and decoder of the question generation model have the same configuration as a BERT-base model.
In addition, we share parameters between encoder and decoder, and parameter values are initialized with the public uncased BERT-base checkpoint\footnote{\scriptsize{\url{https://github.com/google-research/bert}}}.  
We truncate answer passage to 128 tokens and limit decoding to 64 steps. 
The training objective is the standard cross entropy, and the model is trained with a batch size of 128 and learning rate 1e-4 using Adam \citep{adam} optimizer. 

The dual encoder model described in Section~\ref{sec:de} is based on a customized BERT model, which contains 12 transformer layers \citep{transformer}, each layer with 1024 hidden dimension and 16 attention heads.
We pretrain our own BERT model on PubMed abstracts with a customized wordpiece vocabulary which contains 107K entries. 
We follow the same sentence sampling procedure as reported in the original BERT paper, e.g., the combined sequence has length no longer than 512 tokens, and we uni-formly mask $15\%$ of the tokens from each sequence for masked language model prediction.  
We  update  the  next  sentence  prediction  task  by  replacing  original binary-cross-entropy  loss  with  softmax  cross-entropy loss. 
We use the same hyper-parameter values for BERT pretraining as \citet{bert} except that the learning rate is set 2e-5, and the model is trained for 300,000 steps.

For dual encoder training, we use a batch size of 6144.
Each training example in the batch is a question-abstract pair, and we truncate queries and abstracts to 48 and to 350 tokens by BERT wordpiece tokenization.
We train the model for 100,000 steps using Adam with a learning rate 5e-6.
Similar to BERT pretraining, we also apply L2 weight decay of 0.01, and warm up learning rate for the  first 10,000 steps.

\subsection{TFR-BERT Rankers}
\label{sec:systems:bert}

We base our re-ranking strategies on TFR-BERT~\citep{Han+al:2020}.
In general, we fine-tune a pre-trained contextual representation model like $\Bert$ based on ranking losses and score each document $\doc$ (since BERT-style encoders are usually applied to shorter text sequences, in practice we only apply them for scoring document abstracts). Then we re-rank all the documents based on the ranking scores.
We fine-tune the re-ranker model based on different pre-trained models and with different strategies and include these runs into the hierarchical reciprocal rank fusion.


First, we briefly introduce the model structure of the re-ranker. 
For each query $q$ and a candidate document $\doc$ retrieved by the retrieval system, we construct the input sequence of tokens by concatenating the query tokens and the document tokens, separated by \SEP~tokens. 
We also add a \CLS~token at the beginning of the sequence. 
Then, we feed the sequence into an encoder based on pre-trained \Bert-like model and take the output embedding of the \CLS~token.
Take \Bert~as an example, we denote the output embedding as $\be_{\Bert}(\doc)$.
Based on the embedding, we simply use a dense layer to get the ranking score of document $\doc$ by:
\begin{align}
    \score{\Bert}{\doc} = \sigma \big(\bW \be_{\Bert}(\doc) + b \big)
\end{align}
where $\bW$ and $b$ are trainable parameters of the dense layer. 

The entire scorer can be trained with ranking losses for optimal ranking performance.
In this work, we apply a softmax ranking loss. 
For each query $q$, if we denote the retrieved candidate document set as $\corpus$ and the ground-truth relevance of each document $\doc \in \corpus$ as $y_{\doc}$, 
then the softmax ranking loss for this query can be written as:
\begin{align}
    \ell_q  = \sum_{\doc \in \corpus} \frac{y_{\doc}}{\sum_{\doc' \in \corpus} {y_{\doc'}}} \log \Bigg( \frac{\exp(\score{\Bert}{\doc})}{\sum_{\doc' \in \corpus} \exp(\score{\Bert}{\doc'})} \Bigg)
\end{align}
We can fine-tune the entire ranking model by using the softmax ranking function. 
Notice that this loss function is a simplified version of the ranking loss proposed by~\cite{Xia+al:2008}.

It is worth pointing out that the encoder does not need to be pre-trained \Bert. 
There are many similar encoders publicly available with similar structure that can be plugged in seamlessly. 
After trying multiple alternatives, we find that \Electra~\citep{Clark+al:2019} and \Roberta~\citep{Liu+al:2019} are the most effective ones. 
We denote the ranking scorers based on these two encoders as $\score{\Electra}{\cdot}$ and $\score{\Roberta}{\cdot}$ respectively.

\subsubsection{TFR-BERT fine-tuned on MS-MARCO}
\label{sec:systems:bert:msmarco}

Since the relevance labels provided in TREC-COVID dataset are extremely limited, 
we experiment with utilizing external datasets to fine-tune the re-ranking models. 
The MS-MARCO dataset~\citep{Bajaj+al:2016} is a passage ranking dataset which aims to rank passages based on their relevance to questions. 
The dataset contains about 1 million queries and more than 8 million passages. 
For each query, some relevant passages are annotated. 

We fine-tune the re-ranking scorer based on the labeled data in MS-MARCO dataset.
For each query $q$, we first retrieve all the candidate passages $\corpus$ from the retriever. 
Then for each passage $d \in \corpus$ relevant to the query (i.e., $y_d > 0$), we randomly sample another $(l - 1)$ negative passages with $y_{d'} = 0$ and assemble them together as a candidate subset $\corpus' \subset \corpus$ with size $l$. 
We train the re-ranking scorer based on $\corpus'$.
This step reduces the computational requirements and avoid numerical instability instead of feeding more than 1,000 passages into the re-ranker for fine-tuning. 
Notice that although we only sample a very small subset for fine-tuning, it is not necessary for inference. 
Since our re-ranker only needs a query-document pair as input during inference, we can always score all the candidate documents retrieved in the first-stage for each query and re-rank all of them to ensure recall. 

For inference on TREC-COVID dataset, we take the ``query'' and ``question'' segment and directly concatenate them together as the query tokens. We then concatenate \CLS, the query tokens and the passage tokens separated by \SEP~as described above and feed the whole sequence into the fine-tuned scorer. 

We use \Bert-Large pre-trained with whole-word masking, \Electra, and \Roberta~as the encoder respectively. They are fine-tuned on MS-MARCO dataset for 200,000 steps with learning rate 1e-5. The batch size is set to 32 and the candidate subset size is set to $l=12$. The maximum sequence length is set to 512 and any passages resulting in longer sequences will be truncated. 
We keep other fine-tuning configurations such as optimizer and warming-up steps the same as the default \Bert~fine-tuning configurations. 
We fine-tuned 10 individual re-rankers for each encoder type, resulting in 30 re-rankers. 
All of the 30 re-rankers are regarded as a single system \system~and fused together.

\subsubsection{TFR-BERT fine-tuned on TREC-COVID}

As the number of relevance judgments available per topic grew substantially in the final rounds (see Figure~\ref{fig:corpus-growth}(b)), we also attempted to leverage the limited relevance labels from TREC-COVID dataset. 
The overall model structure is the same as before. 
However, since there is only a small number of relevance labels, the re-ranker could quickly overfit if being fine-tuned for too many steps. 
To explore the best number of fine-tuning steps, we randomly sample 20\% of queries from labeled data as validataion dataset and fine-tune the re-ranker on the other 80\% of the dataset. 
We monitor the performance curve on validation dataset and manually select a reasonable number of fine-tuning steps. 
We then fine-tune the re-rankers on all labeled data for the selected number of fine-tuning steps.
Depending on different encoders, the selected number of fine-tuning step vary from 3,000 to 10,000.

Similarly to MS-MARCO, we use \Bert-Large pre-trained with whole-word masking, \Electra, and \Roberta~as the encoders, respectively. The learning rate is still 1e-5 and the batch size is still 32. Due to limited labeled data, we only set the candidate subset size to 6. The maximum sequence length is still 512. 

We also try two different ways to construct the query sequence: 1) concatenating the ``query'' and ``question'' fields as the query sequence; 2) concatenating the ``question'' and ``narrative'' fields as the query sequence. For each query sequence construction method, We fine-tune 3 individual re-rankers for all 3 types of encoders, resulting in 18 re-rankers to be fused together.

\begin{table}[]
    \centering
    \begin{tabular}{|l|c|c|c|c|c|}
    \hline
     & \#Runs fused & NDCG@20 & P@20 & MAP & Recall@1000 \\
     \hline
    RRF(\underline{T}errier) & $14$ & $50.18^\ast$ & $55.56^\ast$ & $23.10^\ast$ & $53.45^\ast$ \\
    RRF(\underline{A}nserini) & $8$ & $51.83^\ast$ & $56.67^\ast$ & $23.68^\ast$ & $54.52^\ast$ \\
    RRF(\underline{D}ual Encoder) & $24$ & $50.55^\ast$ & $56.11^\ast$ & $20.12^\ast$ & $47.51^\ast$  \\
    \hline
    RRF(\underline{TA}) & $26$ & $53.52^\ast$ & $57.78^\ast$ & $25.95^\ast$ & $56.72^\ast$  \\
    RRF(\underline{TAD}) & $50$ & $55.47$ & $60.67$ & $27.44^\ast$ & $59.00$  \\
    \hline
    $h$-RRF(\underline{TAD}) & $3$ & $\mathbf{56.64}$ & $\mathbf{62.22}$ & $\mathbf{27.98}$ & $\mathbf{59.67}$ \\
    \hline
    \end{tabular}
        \caption{Ablation study of the retrieval systems. Individual runs performance is not reported, since we found them to be generally well below the performance of the RRF runs. Statistically significant differences (paired t-test, $p < 0.05$) from the last row are marked by $^\ast$. Best overall metric is bolded.}
    \label{tab:base-ret}
\end{table}

\begin{table}[]
    \centering
    \begin{tabular}{|l|c|c|c|c|c|}
    \hline
    & \#Runs fused & NDCG@20 & P@20 & MAP & Recall@1000 \\
     \hline
    RRF(\underline{M}S-Marco BERT) & $30$ & $52.74^\ast$ & $55.44^\ast$  & $24.11^\ast$ & $56.30^\ast$ \\
    RRF(\underline{F}inetuned BERT) & $18$ & $68.92$ & $70.78$ & $36.36^\ast$ & $67.75^\ast$ \\
    RRF(\underline{R}elevance Feedback) & $4$ & $62.67^\ast$  & $66.56^\ast$  & $30.20^\ast$ & $60.60^\ast$ \\
    \hline
    RRF(\underline{TADM}) & $80$  & $59.97^\ast$ & $64.22^\ast$ & $29.31^\ast$ & $60.86^\ast$ \\
    RRF(\underline{TADMF}) & $98$ & $64.51^\ast$ & $67.11^\ast$  & $33.66^\ast$ & $65.66^\ast$ \\
    RRF(\underline{TADMFR})& $102$ & $64.88^\ast$ & $67.56^\ast$ & $34.07^\ast$ & $65.77^\ast$ \\
    \hline
    $h$-RRF(\underline{TADMFR}) & $6$ & $62.67^\ast$ & $66.56^\ast$ & $30.20^\ast$ & $60.60^\ast$ \\
    $h_w$-RRF(\underline{TADMFR}) & $6$ & $\mathbf{71.61}$ & $\mathbf{72.56}$ & $\mathbf{39.13}$ & $\mathbf{69.36}$ \\
    \hline
    \end{tabular}
    \caption{Ablation study of all the retrieval and ranking runs that comprise the final weighted hierarchical rank fusion ensemble. Individual runs performance is not reported, since we found them to be generally well below the performance of the RRF runs. Statistically significant differences (paired t-test, $p < 0.05$) from the last row are marked by $^\ast$. Best overall metric is bolded.}
    \label{tab:final-ret}
\end{table}

\section{Experimental Results}
\label{sec:exp}

We begin this section by reporting the results of the ablation studies designed to evaluate the various aspects of our overall ensemble approach using relevance judgments from Round 1 -- 4. These analyses were done in a lead up to Round 5, and form the basis for our final submission to this round. Then, we report the official metrics for our best automatic and feedback runs for Round 5 of the TREC-COVID challenge.\footnote{Our submissions performed equally well in Round 4 of the competition, but since these submissions do not neatly correspond to the ensembling approach discussed in this paper, we only report Round 5 results here.}

\subsection{Ablation studies}

In these ablation studies, we use the relevance judgments from Rounds 1 -- 4 to better understand the contributions of the systems to be used in our final ensemble. In Table \ref{tab:base-ret}, we look at the performance of each of the retrieval systems, both lexical (\underline{T}errier and \underline{A}nserini) and semantic (the \underline{D}ual Encoder described in Section \ref{sec:systems:semantic}).

Overall, it is clear from Table \ref{tab:base-ret} that while the three retrieval systems are comparable in their performance (with system \underline{D} slightly trailing the lexical retrieval systems in MAP and Recall@1000), their results are highly complementary. RRF(\underline{TA}) achieves large gains as compared to either Terrier or Anserini. Fusion with Dual Encoder system leads to additional gains, especially in the $h$-RRF(\underline{TAD}) variant, which has a 8\% increase in MAP compared to RRF(\underline{TA}). Overall, $h$-RRF(\underline{TAD}) is statistically significantly better than the other alternatives on most of the reported metrics.

In Table \ref{tab:final-ret}, we switch our attention to the final combination of the retrieval systems with the ranking systems:  \underline{M}S-Marco BERT, \underline{F}inetuned BERT and \underline{R}elevance Feedback. For \underline{R}elevance Feedback, as described in Section~\ref{sec:systems:rf}, we fuse two Terrier relevance feedback retrieval runs, and two Anserini relevance feedback runs provided by the \teamname{covidex} team. For $\text{RRF}(\ast\text{ BERT})$ runs, we use a fusion of multiple rerankers (described in Section~\ref{sec:systems:bert}) each applied to the top $2000$ results from the $h$-RRF(\underline{TAD}) run, the best performing run in Table~\ref{tab:base-ret}.  

In Table~\ref{tab:final-ret}, again, we see a clear indication of the \emph{more is more} principle: ensembles with a larger number of runs achieve better performance. The best unweighted retrieval and ranking ensemble,  RRF(\underline{TADMFR}), achieves an almost 20\% gain in MAP, as compared to the best retrieval-only ensemble, $h$-RRF(\underline{TAD}).

Heuristic weighting of the runs that have access to relevance judgments, as described in Section \ref{sec:ensemble:hwrrf}, results in an additional significant improvement. $h_w$-RRF(\underline{TADMFR}) achieves roughly 15\% and 10\% improvement over the best unweighted ensemble in terms of MAP and NDCG@20, respectively. In both cases these improvements are statistically significant.

With these ablation studies in mind, we use our 102-run weighted hierarchical rank fusion ensemble (a.k.a \teamname{RRF102}) as the highest priority submission for Round 5 of the TREC-COVID challenge. As we were allowed to submit additional runs, we submit other alternative ensemble combinations as well.

\begin{table}[]
    \centering
    \begin{tabular}{|l|c|c|c|c|}
    \hline
    Run ID & NDCG@20 & P@20 & MAP & Recall@1000 \\
     \hline
    \teamname{rk\_ir\_trf\_logit\_rr} & $79.56^\ast$ & $82.60^\ast$  & $37.89^\ast$ & $62.91^\ast$ \\
    \teamname{covidex.r5.2s.lr} & $83.11$ & $84.60$  &  $39.22^\ast$ & $61.47^\ast$ \\
    \teamname{sab20.5.4.dfo} & $77.91^\ast$ & $82.10^\ast$  & $40.61^\ast$ & $72.17^\ast$ \\
    \teamname{elhuyar\_rrf\_nof09p} & $77.89^\ast$ & $83.10^\ast$  & $41.69^\ast$ & $70.68^\ast$ \\
    \hline
    \teamname{UPrrf102-r5} & $80.92^\ast$ & $85.30$  & $45.69^\ast$ & $\mathbf{76.09}^\ast$ \\
    \teamname{UPrrf102-wt-r5} & $\mathbf{84.90}$ & $\mathbf{86.90}$  & $\mathbf{47.31}$ & $75.53$ \\
    \hline
    \end{tabular}
    \caption{Comparison to \emph{feedback} runs by four other top-performing (as measured by NDCG@20 and MAP) teams in TREC-COVID Round 5. The best run per team is used. Runs are sorted by the MAP metric, and statistically significant differences (paired t-test, $p < 0.05$) from the last row are marked by $^\ast$. Best overall metric is bolded.}
    \label{tab:rnd5-fb}
\end{table}

\begin{table}[]
    \centering
    \begin{tabular}{|l|c|c|c|c|}
    \hline
    Run ID & NDCG@20 & P@20 & MAP & Recall@1000 \\
     \hline
    \teamname{covidex.r5.d2q.2s} & $\mathbf{75.39}$ & $77.00$  & $32.27^\ast$ & $60.22^\ast$ \\
    \teamname{uogTrDPH\_QE\_SB\_CB} & $74.27$ & $\mathbf{79.10}$  & $33.05^\ast$ & $59.05^\ast$ \\
    \hline
    \teamname{UPrrf80-r5} & $71.16$ & $75.90$ & $35.98$ & $69.43$ \\
    \teamname{UPrrf89-r5} & $72.35$ & $75.90$ & $\mathbf{36.12}$ & $\mathbf{69.48}$ \\
    \hline
    \end{tabular}
    \caption{Comparison to \emph{automatic} runs by two other top-performing (as measured by NDCG@20 and MAP) teams in TREC-COVID Round 5. The best run per team is used. Runs are sorted by the MAP metric, and statistically significant differences (paired t-test, $p < 0.05$) from the last row are marked by $^\ast$. Best overall metric is bolded.}
    \label{tab:rnd5-auto}
\end{table}

\subsection{TREC-COVID Round 5 Official Results}

In this section, we briefly summarize the official performance of our runs in Round 5 of the challenge. Since TREC-COVID challenge uses residual collection evaluation, all the documents that were evaluated in Rounds 1 -- 4 are filtered out from the submitted runs.

Table~\ref{tab:rnd5-fb} compares the performance of the weighted hierarchical reciprocal rank fusion run \teamname{UPrrf102-wt-r5} (equivalent to $h_w$-RRF(\underline{TADMFR}) in Table~\ref{tab:final-ret}) to four other runs by top ranked teams, as well as our \emph{unweighted} variant \teamname{UPrrf102-r5} (equivalent to $h$-RRF(\underline{TADMFR}) in Table~\ref{tab:final-ret}). \teamname{UPrrf102-wt-r5} outperforms all other submissions, in most cases to a statistically significant degree. In particular, the increases in MAP and Recall@1000 are especially impressive. \teamname{UPrrf102-wt-r5} achieves $13.4\%$ MAP gain as compared to the next best team's run (\teamname{elhuyar\_rrf\_nof09p}). This demonstrates the utmost importance of retrieval and ranking ensembles for systems that require high relevant document recall.

In addition to \emph{feedback} runs, i.e., runs that are produced using systems that have access to relevance labels from prior rounds, TREC-COVID challenge allowed submission of automatic runs -- runs that were not tuned or modified using prior relevance judgments. We submitted two such runs, \teamname{UPrrf80-r5} and \teamname{UPrrf89-r5}, that are compared to other top-performing automatic runs in Table~\ref{tab:rnd5-auto}.

\teamname{UPrrf80-r5} is equivalent to RRF(\underline{TADM}) in Table~\ref{tab:final-ret}, which fuses Terrier, Anserini, Dual Encoder and MS-Marco BERT system runs. \teamname{UPrrf89-r5} incorporates 9 additional runs fine-tuned on BioASQ\footnote{\scriptsize{\url{http://www.bioasq.org/}}}, a document ranking task dataset with biomedical questions.
We use questions from year 1 to 5 of the BioASQ competition.
We follow the same data split as \citet{McDonald2018DeepRR}, where we use year 1 to 4 as training data, 
and use batch 1 of year 5 for tuning.  
Negative passages are abstracts of documents returned by a BM25 system.
The BioASQ re-ranker is fine-tuned almost in the same manner as described in Section~\ref{sec:systems:bert:msmarco} for MS-MARCO re-ranker, except that we set the candidate subset size as $l=6$ and the number of fine-tuning steps to 10,000. Neither \teamname{UPrrf80-r5} nor \teamname{UPrrf89-r5} use any information from TREC-COVID, and thus can be classified as automatic runs.

Overall, these automatic runs once again demonstrate the importance of retrieval and ranking ensembles for achieving high recall. \teamname{UPrrf89-r5} achieves $9.2\%$ MAP gain as compared to the next best team's run (\teamname{uogTrDPH\_QE\_SB\_CB}). While our runs are not ranked the highest in terms of NDCG@20 and P@20 metrics, the difference from the top runs by \texttt{covidex} and \texttt{uogTr} were not found to be statistically significant in our analysis. In addition, while \teamname{UPrrf89-r5} slightly outperforms \teamname{UPrrf80-r5} on all metrics, no statistically significant differences were found between the two runs.

\section{Conclusions}
The TREC-COVID challenge organizers brought to the forefront the importance of delivering accurate, reliable, complete and up-to-date information to scientists, medical practitioners, and government officials in the midst of rapidly evolving pandemic. To meet this challenge, in this paper, we describe a weighted hierarchical rank fusion ensemble approach that synthesizes 102 runs from lexical and semantic retrieval systems, pre-trained and fine-tuned BERT rankers, as well as relevance feedback runs. We hypothesize that such an ensemble can effectively leverage the complementary nature of its constituents, and provide a high recall of relevant documents to the searcher, while maintaining high precision at top ranks.

The proposed ensemble achieves state-of-the-art performance in rounds 4 and 5 of the TREC-COVID challenge, outperforming submissions by 27 other teams. While the approach described here is conceptually simple, we found it to be highly robust to collection dynamics, as well as being effective at achieving state-of-the art performance on a wide range of metrics, including NDCG and precision at top ranks, mean average precision, and relevant document recall. In future work, we plan to explore further improvements to this approach, as well as its applications to settings beyond biomedical search.

\section*{Acknowledgments}
We thank the TREC-COVID challenge organizers for their tremendous investment of time and effort in smooth execution of all the evaluation rounds. We thank our fellow challenge participants for insightful discussions on the \texttt{trec-covid} forum. In particular, we thank Jimmy Lin and the \texttt{covidex} teams for publishing their Anserini runs. This work would not be possible without the support provided by the TF-Ranking team.

\newpage

\bibliographystyle{unsrtnat}
\bibliography{trec-covid}

\begin{thebibliography}{26}
\providecommand{\natexlab}[1]{#1}
\providecommand{\url}[1]{\texttt{#1}}
\expandafter\ifx\csname urlstyle\endcsname\relax
  \providecommand{\doi}[1]{doi: #1}\else
  \providecommand{\doi}{doi: \begingroup \urlstyle{rm}\Url}\fi

\bibitem[Voorhees et~al.(2020)Voorhees, Alam, Bedrick, Demner-Fushman, Hersh,
  Lo, Roberts, Soboroff, and Wang]{Voorhees+al:2020}
Ellen Voorhees, Tasmeer Alam, Steven Bedrick, Dina Demner-Fushman, William~R
  Hersh, Kyle Lo, Kirk Roberts, Ian Soboroff, and Lucy~Lu Wang.
\newblock Trec-covid: Constructing a pandemic information retrieval test
  collection.
\newblock \emph{arXiv preprint arXiv:2005.04474}, 2020.

\bibitem[Zhang et~al.(2020)Zhang, Gupta, Tang, Han, Pradeep, Lu, Zhang,
  Nogueira, Cho, Fang, and Lin]{Zhang+al:2020}
Edwin Zhang, Nikhil Gupta, Raphael Tang, Xiao Han, Ronak Pradeep, Kuang Lu, Yue
  Zhang, Rodrigo Nogueira, Kyunghyun Cho, Hui Fang, and Jimmy Lin.
\newblock Covidex: Neural ranking models and keyword search infrastructure for
  the covid-19 open research dataset.
\newblock \emph{arXiv preprint arXiv:2007.07846}, 2020.

\bibitem[MacAvaney et~al.(2020)MacAvaney, Cohan, and
  Goharian]{MacAvaney+al:2020}
Sean MacAvaney, Arman Cohan, and Nazli Goharian.
\newblock Sledge: A simple yet effective baseline for covid-19 scientific
  knowledge search.
\newblock \emph{arXiv preprint arXiv:2005.02365}, 2020.

\bibitem[Bell and Koren(2007)]{Bell+Koren:2007}
Robert~M. Bell and Yehuda Koren.
\newblock Lessons from the netflix prize challenge.
\newblock \emph{ACM SIGKDD Explorations Newsletter}, 9\penalty0 (2):\penalty0
  75–79, December 2007.

\bibitem[Burges et~al.(2011)Burges, Svore, Bennett, Pastusiak, and
  Wu]{Burges+al:2011}
Christopher Burges, Krysta Svore, Paul Bennett, Andrzej Pastusiak, and Qiang
  Wu.
\newblock Learning to rank using an ensemble of lambda-gradient models.
\newblock In \emph{Proceedings of the learning to rank Challenge}, pages
  25--35, 2011.

\bibitem[Han et~al.(2020)Han, Wang, Bendersky, and Najork]{Han+al:2020}
Shuguang Han, Xuanhui Wang, Mike Bendersky, and Marc Najork.
\newblock Learning-to-rank with bert in tf-ranking.
\newblock \emph{arXiv preprint arXiv:2004.08476}, 2020.

\bibitem[Hastie et~al.(2009)Hastie, Tibshirani, and Friedman]{Hastie+al:2009}
Trevor Hastie, Robert Tibshirani, and Jerome Friedman.
\newblock \emph{The elements of statistical learning: data mining, inference,
  and prediction}.
\newblock Springer Science \& Business Media, 2009.

\bibitem[Kanoulas et~al.(2017)Kanoulas, Li, Azzopardi, and
  Spijker]{Kanoulas+al:2017}
Evangelos Kanoulas, Dan Li, Leif Azzopardi, and Rene Spijker.
\newblock Clef 2017 technologically assisted reviews in empirical medicine
  overview.
\newblock In \emph{CEUR Workshop Proceedings}, volume 1866, pages 1--29, 2017.

\bibitem[Cormack et~al.(2009)Cormack, Clarke, and Buettcher]{Cormack+al:2009}
Gordon~V. Cormack, Charles L~A Clarke, and Stefan Buettcher.
\newblock Reciprocal rank fusion outperforms condorcet and individual rank
  learning methods.
\newblock In \emph{Proceedings of SIGIR}, page 758–759, 2009.

\bibitem[Ounis et~al.(2005)Ounis, Amati, Plachouras, He, Macdonald, and
  Johnson]{Terrier:2005}
Iadh Ounis, Gianni Amati, Vassilis Plachouras, Ben He, Craig Macdonald, and
  Douglas Johnson.
\newblock Terrier information retrieval platform.
\newblock In \emph{European Conference on Information Retrieval}, pages
  517--519. Springer, 2005.

\bibitem[Yang et~al.(2017)Yang, Fang, and Lin]{Anserini:2017}
Peilin Yang, Hui Fang, and Jimmy Lin.
\newblock Anserini: Enabling the use of lucene for information retrieval
  research.
\newblock In \emph{Proceedings of SIGIR}, page 1253–1256, 2017.

\bibitem[Peng et~al.(2007)Peng, Macdonald, He, Plachouras, and
  Ounis]{Peng+al:2007}
Jie Peng, Craig Macdonald, Ben He, Vassilis Plachouras, and Iadh Ounis.
\newblock Incorporating term dependency in the dfr framework.
\newblock In \emph{Proceedings of SIGIR}, pages 843--844, 2007.

\bibitem[Palangi et~al.(2016)Palangi, Deng, Shen, Gao, He, Chen, Song, and
  Ward]{palangi2016deep}
Hamid Palangi, Li~Deng, Yelong Shen, Jianfeng Gao, Xiaodong He, Jianshu Chen,
  Xinying Song, and Rabab Ward.
\newblock Deep sentence embedding using long short-term memory networks:
  Analysis and application to information retrieval.
\newblock \emph{IEEE/ACM Transactions on Audio, Speech, and Language
  Processing}, 24\penalty0 (4):\penalty0 694--707, 2016.

\bibitem[Devlin et~al.(2019)Devlin, Chang, Lee, and Toutanova]{bert}
Jacob Devlin, Ming-Wei Chang, Kenton Lee, and Kristina Toutanova.
\newblock {BERT}: Pre-training of deep bidirectional transformers for language
  understanding.
\newblock In \emph{Proceedings of NAACL}, June 2019.

\bibitem[Gillick et~al.(2018)Gillick, Presta, and Tomar]{GillickE2E}
Daniel Gillick, Alessandro Presta, and Gaurav~Singh Tomar.
\newblock End-to-end retrieval in continuous space.
\newblock \emph{arXiv preprint arXiv:1811.08008}, 2018.

\bibitem[Karpukhin et~al.(2020)Karpukhin, Oğuz, Min, Wu, Edunov, Chen, and
  Yih]{dense-passage-qa}
Vladimir Karpukhin, Barlas Oğuz, Sewon Min, Ledell Wu, Sergey Edunov, Danqi
  Chen, and Wen-tau Yih.
\newblock Dense passage retrieval for open-domain question answering.
\newblock \emph{arXiv preprint arXiv:2004.04906}, 2020.

\bibitem[Liu et~al.(2011)Liu, Wang, Kumar, and Chang]{liu2011hashing}
Wei Liu, Jun Wang, Sanjiv Kumar, and Shih-Fu Chang.
\newblock Hashing with graphs.
\newblock In \emph{Proceedings of the International Conference on Machine
  Learning}, 2011.

\bibitem[Johnson et~al.(2017)Johnson, Douze, and J{\'e}gou]{JDH17}
Jeff Johnson, Matthijs Douze, and Herv{\'e} J{\'e}gou.
\newblock Billion-scale similarity search with gpus.
\newblock \emph{arXiv preprint arXiv:1702.08734}, 2017.

\bibitem[Ma et~al.(2020)Ma, Korotkov, Yang, Hall, and
  McDonald]{Ma2020ZeroshotNR}
Ji~Ma, I.~Korotkov, Yin-Fei Yang, K.~Hall, and R.~McDonald.
\newblock Zero-shot neural retrieval via domain-targeted synthetic query
  generation.
\newblock \emph{arXiv preprint arXiv:2004.14503}, 2020.

\bibitem[Vaswani et~al.(2017)Vaswani, Shazeer, Parmar, Uszkoreit, Jones, Gomez,
  Kaiser, and Polosukhin]{transformer}
Ashish Vaswani, Noam Shazeer, Niki Parmar, Jakob Uszkoreit, Llion Jones,
  Aidan~N Gomez, \L~ukasz Kaiser, and Illia Polosukhin.
\newblock Attention is all you need.
\newblock In \emph{Advances in Neural Information Processing Systems}, pages
  5998--6008. 2017.

\bibitem[Kingma and Ba(2014)]{adam}
Diederik Kingma and Jimmy Ba.
\newblock Adam: A method for stochastic optimization.
\newblock In \emph{Proceedings of ICLR}, 2014.

\bibitem[Xia et~al.(2008)Xia, Liu, Wang, Zhang, and Li]{Xia+al:2008}
Fen Xia, Tie-Yan Liu, Jue Wang, Wensheng Zhang, and Hang Li.
\newblock Listwise approach to learning to rank: theory and algorithm.
\newblock In \emph{Proceedings of the 25th international conference on Machine
  learning}, pages 1192--1199, 2008.

\bibitem[Clark et~al.(2019)Clark, Luong, Le, and Manning]{Clark+al:2019}
Kevin Clark, Minh-Thang Luong, Quoc~V Le, and Christopher~D Manning.
\newblock Electra: Pre-training text encoders as discriminators rather than
  generators.
\newblock In \emph{Proceedings of ICLR}, 2019.

\bibitem[Liu et~al.(2019)Liu, Ott, Goyal, Du, Joshi, Chen, Levy, Lewis,
  Zettlemoyer, and Stoyanov]{Liu+al:2019}
Yinhan Liu, Myle Ott, Naman Goyal, Jingfei Du, Mandar Joshi, Danqi Chen, Omer
  Levy, Mike Lewis, Luke Zettlemoyer, and Veselin Stoyanov.
\newblock Roberta: A robustly optimized bert pretraining approach.
\newblock \emph{arXiv preprint arXiv:1907.11692}, 2019.

\bibitem[Bajaj et~al.(2016)Bajaj, Campos, Craswell, Deng, Gao, Liu, Majumder,
  McNamara, Mitra, Nguyen, et~al.]{Bajaj+al:2016}
Payal Bajaj, Daniel Campos, Nick Craswell, Li~Deng, Jianfeng Gao, Xiaodong Liu,
  Rangan Majumder, Andrew McNamara, Bhaskar Mitra, Tri Nguyen, et~al.
\newblock Ms marco: A human generated machine reading comprehension dataset.
\newblock \emph{arXiv preprint arXiv:1611.09268}, 2016.

\bibitem[McDonald et~al.(2018)McDonald, Brokos, and
  Androutsopoulos]{McDonald2018DeepRR}
Ryan McDonald, George Brokos, and Ion Androutsopoulos.
\newblock Deep relevance ranking using enhanced document-query interactions.
\newblock In \emph{Proceedings of EMNLP}, 2018.

\end{thebibliography}
\end{document}